\documentclass[12pt]{article}
\usepackage{amsmath}
\usepackage{amssymb}
\usepackage{fancyhdr}
\usepackage{graphicx}

\renewcommand{\maketitle}{
    \begin{center}
      \large
        {\bf Discrete Symmetries of Off-Shell Electromagnetism}
        \vskip .3 true cm
      \normalsize
        Martin Land \\
        \vskip .3 true cm
        Department of Computer Science \\
        Hadassah College \\
        P. O. Box 1114, Jerusalem 91010, Israel \\
	email: martin@multinet.net.il
      \end{center}
      \vskip .3 true cm
}

\begin{document}

\title{}
\author{}
\maketitle

\begin{abstract}
This paper discusses the discrete symmetries of off-shell electromagnetism,
the Stueckelberg-Schrodinger relativistic quantum theory and its associated
5D local gauge theory. Seeking a dynamical description of
particle/antiparticle interactions, Stueckelberg developed a covariant
mechanics with a monotonically increasing Poincar\'{e}-invariant parameter.
In Stueckelberg's framework, worldlines are traced out through the
parameterized evolution of spacetime events, which may advance or retreat
with respect to the laboratory clock, depending on the sign of the energy,
so that negative energy trajectories appear as antiparticles when the
observer describes the evolution using the laboratory clock. The associated
gauge theory describes local interactions between events (correlated by the
invariant parameter) mediated by five off-shell gauge fields. These gauge
fields are shown to transform tensorially under under space and time
reflections --- unlike the standard Maxwell fields --- and the interacting
quantum theory therefore remains manifestly Lorentz covariant. Charge
conjugation symmetry in the quantum theory is achieved by simultaneous
reflection of the sense of evolution and the fifth scalar field. Applying
this procedure to the classical gauge theory leads to a purely classical
manifestation of charge conjugation, placing the $CPT$ symmetries on the
same footing in the classical and quantum domains. In the resulting picture,
interactions do not distinguish between particle and antiparticle
trajectories --- charge conjugation merely describes the interpretation of
observed negative energy trajectories according to the laboratory clock.
\end{abstract}

\baselineskip7mm \parindent=0cm \parskip=10pt


\section{Introduction}

\subsection{Stueckelberg's Model of Pair Creation/Annihilation}

In 1941, Stueckelberg \cite{Stueckelberg} proposed a covariant Hamiltonian
formalism for interacting spacetime events, in which the events evolve
dynamically, as functions of a Poincar\'{e} invariant parameter (see also
Fock \cite{Fock}). In the classical mechanics, the particle worldline is
traced out in terms of the values taken on by the four-vector $x^{\mu
}\left( \tau \right) $ as the parameter proceeds monotonically from $\tau
=-\infty $ to $\tau =\infty $. Stueckelberg's purpose was to ascribe pair
creation/annihilation to a single worldline, generated dynamically by an
event whose time coordinate advances or retreats with respect to the
laboratory clock, as its instantaneous energy changes sign under interaction
with a field. Figure 1 is a reconstruction of the corresponding illustration
in Stueckelberg's paper \cite{Stueckelberg}.

\begin{center}
\includegraphics[width=250pt]{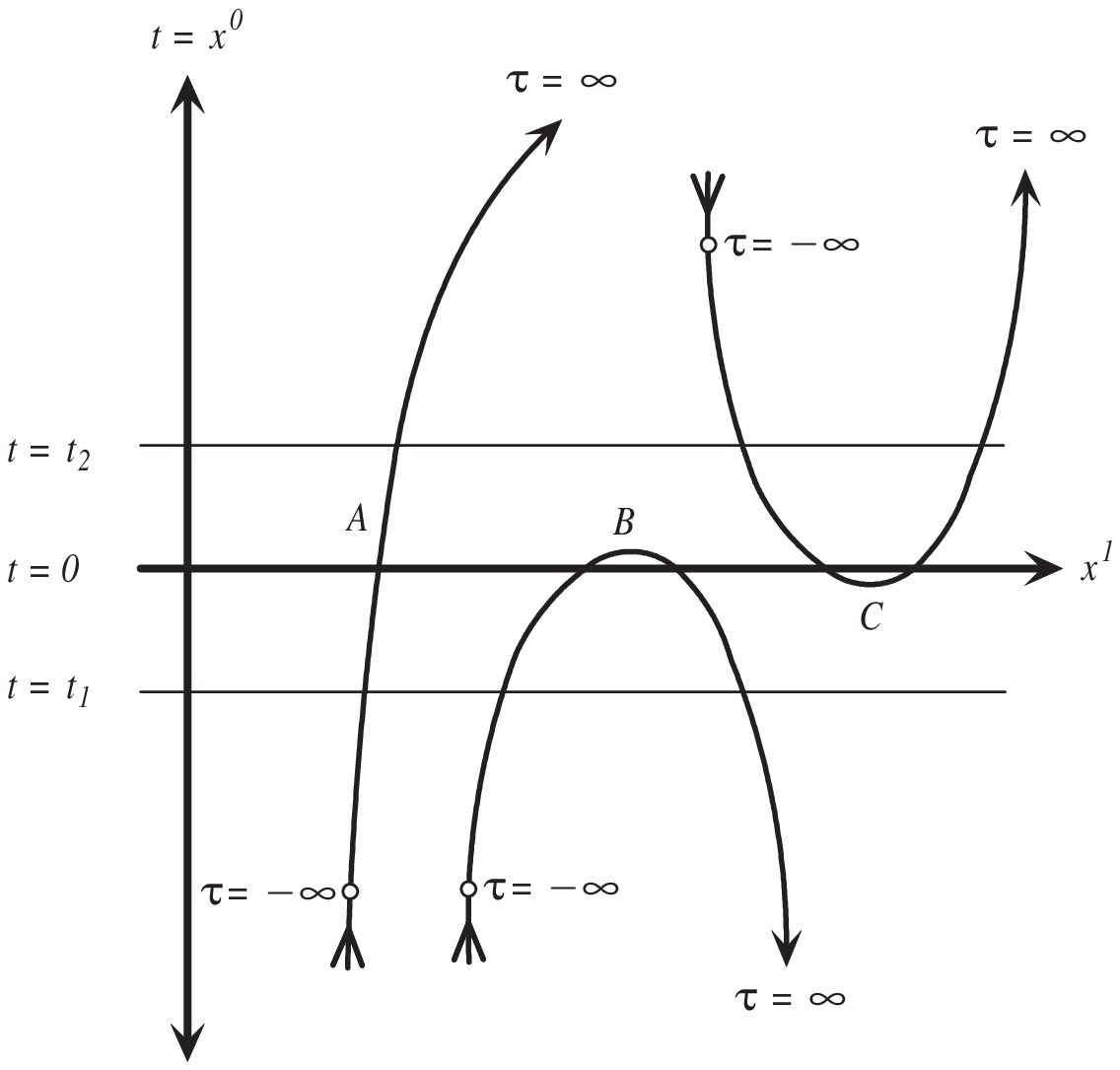}

\begin{tabular}{l}
{\footnotesize \ \textbf{Figure 1}: World Lines} \\ 
\begin{tabular}{l}
{\footnotesize A: Usual type, with a unique solution to }$t\left( \tau
\right) =x^{0}${\footnotesize \ for each }$x^{0}$ \\ 
{\footnotesize B: Annihilation type, with two solutions to }$t\left( \tau
\right) =x^{0}${\footnotesize \ for }$x^{0}\ll 0${\footnotesize \ and no
solution for }$x^{0}\gg 0$ \\ 
{\footnotesize C: Creation type, with two solutions to }$t\left( \tau
\right) =x^{0}${\footnotesize \ for }$x^{0}\gg 0${\footnotesize \ and no
solution for }$x^{0}\ll 0$%
\end{tabular}%
\end{tabular}
\end{center}

The invariant parameter $\tau$ is required because the worldlines
of Figure 1 are generally not single-valued in spacetime.  By
regarding $\tau$ as a physical time parameter, Stueckelberg
recognized two aspects of time \cite{Two-Aspects}, explicitly
distinguishing the Einstein coordinate time $x^{0}$ from the
temporal order $\tau $.  Writing $\tau$-dependent equations of
motion, Stueckelberg was led to a description of the antiparticle
which is similar to Feynman's, but differs in its implementation
of the discrete symmetries.  In the standard treatment of $CPT$,
based on Wigner's approach to time reversal \cite{movies}, the
$T$ operation is understood as both inversion of the time
coordinate and reversal of motion --- inversion of the temporal
ordering of events. Despite the similarity of the worldlines in
Figure 1 to Feynman's spacetime diagrams, the explicit
parameterization of the curves by $\tau $ formally distinguishes
the two aspects of time, and hence two notions of time reversal.
Formal analysis of the dynamical framework arising from
Stueckelberg's approach shows that, unlike the standard $CPT$
implementation, fields and currents transform tensorially under
the discrete Lorentz transformations. The charge conjugation
operation is seen to connect laboratory observation with
experiment, but does not play a role in interactions. Although
$CPT$ invariance was not established as a fundamental symmetry
until more than a decade after Stueckelberg and Feynman's initial
work on this subject, and they did not utilize that approach,
these symmetry properties expose the different interpretations of
the antiparticle.

When standard textbooks (see for example \cite{IZ}) discuss what is
generally known as the Feynman-Stueckelberg approach, which represents
antiparticles as negative energy modes propagating toward earlier times,
they pose this model as an elegant field theoretic replacement for the older
Dirac hole theory. However, Stueckelberg's paper addresses itself to a
slightly different set of concerns. While hole theory predicted
antiparticles in an attempt to solve the problem of negative energy
solutions to the Dirac and Klein-Gordon equations, Stueckelberg's goal was
the formulation of a relativistic generalization of classical and quantum
mechanics\footnote{%
Introducing his generalized Lorentz force, Stueckelberg suggests that the
correct formulation of relativistic dynamics was not yet known, when he
states, ``La question se pose de savoir s'il est possible d'\'etablir une
m\'echanique covariante au sens d'Einstein, qui permette l'existence de
telles courbes.''} capable of generating the curves of Figure 1. By 1941,
pair creation and annihilation were regarded as experimentally observable
phenomena, and the historical reasons they were first predicted (as an
artifact of one or another candidate theory) were not necessarily relevant
to their interpretation. Stueckelberg actually emphasized the negative
energy trajectories as an advantage of his theory, not a problem to be
solved. Since the parameter time $\tau $ is formally similar to the Galilean
invariant time in Newtonian theory, this formalism served Stueckelberg's
wider goal of generalizing the techniques of non-relativistic classical and
quantum mechanics to covariant form.

While Stueckelberg defined classical and quantum states explicitly labeled
by $\tau $, Feynman worked with quantum states, in which events are
temporally ordered by algebraic segregation into initial states and final
states. The Feynman prescription for the propagator is equivalent to Dyson's %
\hbox{$t$-ordered} product, which by exchanging particle creation and
annihilation, can be understood as enforcing temporal ordering of in-events
and out-events at the endpoints of a worldline. In discussing the path
integral for the Klein-Gordon equation, Feynman observed \cite{Feynman} that
explicitly labeling the temporal order of events by $\tau $ and assuming
retarded propagation (with respect to $\tau $), leads to the $t$-ordering
prescription for the propagator. However, we will see that the explicit
labeling by $\tau $ affects the meanings of time reversal symmetry and
charge conjugation.

Stueckelberg argued that pair annihilation is observed in worldlines of type
B in Figure 1, in the fact that there are two solutions to $t(\tau )=t_{1}$,
but no solution to $t(\tau )=t_{2}$. The observer will therefore first
encounter two particle trajectories and then encounter none. That the
magnitude of the electric charge should not change along the single
worldline seems clear enough, but the identification of one part of the
worldline as an antiparticle trajectory, further requires that the charge
reverse sign. While this charge reflection may be grasped intuitively ---
carrying positive charge in one time direction is taken as equivalent to
carrying negative charge in the opposite time direction --- in standard
field theory, charge conjugation is demonstrated through the action of the
charge operator. In the parameterized formalism, the charge inversion
appears directly at the classical level: as the event $x^{\mu }\left( \tau
\right) $ evolves toward earlier values of $t=x^{0}$, the slope $%
dx^{0}/d\tau $ must become negative, and in Stueckelberg's generalized
Lorentz force, this derivative multiplies the electric charge. Thus,
particles and antiparticles do not appear as distinct classes of solutions
to a defining equation, but as a single event whose qualitative behavior
depends instantaneously on the dynamical value of its velocity.

\subsection{Gauge Theory}

To generalize the Lorentz equations, Stueckelberg proposed the covariant
evolution equation 
\begin{equation}
\dfrac{d^{2}x^{\mu }}{d\tau ^{2}}=-\Gamma _{\nu \rho }^{\mu }\dfrac{dx^{\nu }%
}{d\tau }\dfrac{dx^{\rho }}{d\tau }+eF^{\mu \nu }g_{\nu \rho }\dfrac{%
dx^{\rho }}{d\tau }+K^{\mu }  \label{10}
\end{equation}%
in which the metric signature is $\mathrm{diag}(-,+,+,+)$ with index
convention $\mu ,\nu ,\rho =0,\cdots ,3$, $F^{\mu \nu }$ is the
electromagnetic field strength tensor, and $\Gamma _{\nu \rho }^{\mu }$ is
an affine connection expressing the influence of gravitation. Stueckelberg
observed that the mass 
\begin{equation}
m^{2}=-g_{\nu \rho }\dfrac{dx^{\nu }}{d\tau }\dfrac{dx^{\rho }}{d\tau }
\label{30}
\end{equation}%
is a constant of integration for $K^{\mu }=0$. The proper time is found by
scaling the invariant parameter through $ds=\pm \sqrt{ds^{2}}=\pm m\ d\tau $
so that when $K^{\mu }=0$, 
\begin{equation}
-g_{\nu \rho }~\dfrac{dx^{\nu }}{md\tau }~\dfrac{dx^{\rho }}{md\tau }=-%
\dfrac{dx^{\nu }}{ds}\dfrac{dx_{\nu }}{ds}=1\;\ \ .  \label{40}
\end{equation}%
Stueckelberg found no reason to claim the existence of the field $K^{\mu }$,
without which the dynamical conservation of mass prevents the classical
worldlines from entering the spacelike region, required for the transition
from positive to negative energy. In the absence of $K^{\mu }$,
Stueckelberg's equation may be derived from the classical Lagrangian 
\begin{equation}
L=\frac{1}{2}M\dot{x}^{\mu }\dot{x}_{\mu }+e\dot{x}^{\mu }A_{\mu }\left(
x\right)  \label{60}
\end{equation}%
and Euler-Lagrange equations 
\begin{equation}
\dfrac{d}{d\tau }\dfrac{\partial L}{\partial \dot{x}_{\mu }}-\dfrac{\partial
L}{\partial x_{\mu }}=0\;\;,  \label{50}
\end{equation}%
where 
\[
F^{\mu \nu }=\partial ^{\mu }A^{\nu }-\partial ^{\nu }A^{\mu } 
\]%
and we have introduced the constant parameter $M$ with dimension of mass.
The equivalent flat space Hamiltonian formulation 
\begin{equation}
K=\frac{1}{2M}(p^{\mu }-eA^{\mu })(p_{\mu }-eA_{\mu })  \label{70}
\end{equation}%
with symplectic equations 
\begin{equation}
\dfrac{dx^{\mu }}{d\tau }=\dot{x}^{\mu }=\dfrac{\partial K}{\partial p_{\mu }%
}\;\;\ \ \ \ ~\ \ \dfrac{dp^{\mu }}{d\tau }=\;\dot{p}^{\mu }=-\dfrac{%
\partial K}{\partial x_{\mu }}\;\ \ ,  \label{80}
\end{equation}%
leads to a quantum theory defined by the equation, 
\begin{equation}
i\partial _{\tau }\psi (x,\tau )=\frac{1}{2M}(p^{\mu }-eA^{\mu })(p_{\mu
}-eA_{\mu })\psi (x,\tau )\;\;.  \label{90}
\end{equation}%
This quantum theory (see also \cite{Feynman}), enjoys the standard U(1)
gauge invariance under local transformations of the type 
\begin{eqnarray}
\psi (x,\tau ) &\longrightarrow &\exp \left[ ie\Lambda (x)\right] \ \psi
(x,\tau )  \label{100} \\
A_{\mu } &\longrightarrow &A_{\mu }+\partial _{\mu }\Lambda (x)  \label{110}
\end{eqnarray}%
but the global gauge invariance is associated with the \emph{five-dimensional%
} conserved current 
\begin{equation}
\partial _{\mu }j^{\mu }+\partial _{\tau }\rho =0  \label{140}
\end{equation}%
where 
\begin{equation}
\rho =\Bigl|\psi (x,\tau )\Bigr|^{2}\qquad j^{\mu }=-\frac{i}{2M}\Bigl\{\psi
^{\ast }(\partial ^{\mu }-ieA^{\mu })\psi -\psi (\partial ^{\mu }+ieA^{\mu
})\psi ^{\ast }\Bigr\}\ .  \label{150}
\end{equation}%
Stueckelberg \cite{Stueckelberg} regarded (\ref{150}) as a true current,
\smallskip leading to the interpretation of $\Bigl|\psi (x,\tau )\Bigr|^{2}$
as the probability density at $\tau $ of finding the event at the spacetime
point $x$. However, under this interpretation, the non-zero divergence of
the four-vector current $j^{\mu }(x,\tau )$ prevents its identification as
the source of the $A^{\mu }(x)$. As a remedy, Stueckelberg observed that
assuming $\rho \rightarrow 0$ pointwise as $\tau \rightarrow \pm \infty $,
integration of (\ref{150}) over $\tau $ leads to 
\begin{equation}
\partial _{\mu }J^{\mu }=0\qquad \text{where}\qquad J^{\mu
}(x)=\int_{-\infty }^{\infty }d\tau \;j^{\mu }(x,\tau )\;.  \label{160}
\end{equation}%
However, in the resulting dynamical picture, the fields $A^{\mu }(x)$ which
mediate particle interaction instantaneously at $\tau $, are induced by
currents $J^{\mu }(x)$ whose support covers the particle worldlines, past
and future. There is no a priori assurance that the particles moving in
these Maxwell fields will trace out precisely the worldlines which induce
the fields responsible for their motion.

In order to obtain a well-posed theory, Sa'ad, Horwitz, and Arshansky \cite%
{saad} introduced a $\tau $-dependent gauge field (see also \cite{Kypr}) and
a \emph{fifth} gauge compensation field, leading to a theory which differs
in significant aspects from conventional electrodynamics, but whose zero
modes coincide with the Maxwell theory. Writing $x^{5}=\tau $ and adopting
the index convention 
\begin{equation}
\lambda ,\mu ,\nu =0,1,2,3\qquad \qquad \mathrm{and}\qquad \qquad \alpha
,\beta ,\gamma =0,1,2,3,5  \label{220}
\end{equation}%
the Stueckelberg-Schrodinger equation%
\begin{equation}
(i\partial _{\tau }+e_{0}a_{5})\psi (x,\tau )=\frac{1}{2M}(p^{\mu
}-e_{0}a^{\mu })(p_{\mu }-e_{0}a_{\mu })\psi (x,\tau )  \label{170}
\end{equation}%
is invariant under the enlarged set of gauge transformations, 
\begin{eqnarray}
\psi (x,\tau ) &\rightarrow &e^{ie_{0}\Lambda (x,\tau )}\psi (x,\tau )
\label{172} \\
a_{\alpha }(x,\tau ) &\rightarrow &a_{\alpha }(x,\tau )+\partial _{\alpha
}\Lambda (x,\tau )  \label{178}
\end{eqnarray}%
and admits the modified five dimensional conserved current 
\begin{equation}
\partial _{\alpha }j^{\alpha }=\partial _{\mu }j^{\mu }+\partial _{\tau
}j^{5}=0  \label{180}
\end{equation}%
where the probability density interpretation still holds for 
\begin{equation}
j^{5}=\Bigl|\psi (x,\tau )\Bigr|^{2}  \label{200}
\end{equation}%
and the current becomes $\tau $-dependent through both the particle and the
gauge fields 
\begin{equation}
j^{\mu }=\frac{-i}{2M}\Bigl[\psi ^{\ast }(\partial ^{\mu }-ie_{0}a^{\mu
})\psi -\psi (\partial ^{\mu }+ie_{0}a^{\mu })\psi ^{\ast }\Bigr]\;.
\label{210}
\end{equation}

The Stueckelberg-Schrodinger equation (\ref{170}) may be derived by
variation of the action 
\begin{equation}
\mathrm{S}=\int d^{4}xd\tau \left\{ \psi ^{\ast }(i\partial _{\tau
}+e_{0}a_{5})\psi -\frac{1}{2M}\psi ^{\ast }(p_{\mu }-e_{0}a_{\mu })(p^{\mu
}-e_{0}a^{\mu })\psi -\frac{\lambda }{4}f_{\alpha \beta }f^{\alpha \beta
}\right\}  \label{360}
\end{equation}%
which includes a kinetic term for the fields, formed from the gauge
invariant quantity 
\begin{equation}
f_{\alpha \beta }=\partial _{\alpha }a_{\beta }-\partial _{\beta }a_{\alpha
}\;.  \label{370}
\end{equation}%
Sa'ad, et.\ al.\ formally raise the index $\beta =5$ in the term $f_{\mu
5}=\partial _{\mu }a_{5}-\partial _{\tau }a_{\mu }$ with the flat metric%
\begin{equation}
g^{\alpha \beta }=\mathrm{diag}(-1,1,1,1,\sigma )\ \;,\;\;g^{55}=\sigma =\pm
1\;,  \label{390}
\end{equation}%
corresponding to a O(4,1) or O(3,2) symmetry, which must break to O(3,1) in
the presence of currents. Varying the action (\ref{360}) with respect to the
gauge fields, the equations of motion are found to be 
\begin{equation}
\partial _{\beta }f^{\alpha \beta }=\frac{e_{0}}{\lambda }j^{\alpha
}=ej^{\alpha }\ \ \ \ \ \ \ \ \ \ \ \ \ \ \ \ \ \ \ \epsilon ^{\alpha \beta
\gamma \delta \epsilon }\partial _{\alpha }f_{\beta \gamma }=0  \label{400}
\end{equation}%
where $j^{\alpha }$ is given in (\ref{200}) and (\ref{210}). Although $%
\lambda $ and $e_{0}$ must be dimensional constants, the dimensionless ratio 
$e_{0}/\lambda $ is the Maxwell charge $e$. In four-vector component form, (%
\ref{400}) becomes 
\begin{equation}
\partial _{\nu }\;f^{\mu \nu }-\partial _{\tau }\;f^{5\mu }=ej^{\mu }\qquad
\qquad \partial _{\mu }\;f^{5\mu }=e\rho .  \label{420}
\end{equation}%
\begin{equation}
\partial _{\mu }f_{\nu \rho }+\partial _{\nu }f_{\rho \mu }+\partial _{\rho
}f_{\mu \nu }=0\qquad \qquad \partial _{\mu }f_{5\nu }-\partial _{\nu
}f_{5\mu }-\partial _{\tau }f_{\mu \nu }=0\;,  \label{430}
\end{equation}%
which may be seen as a four-dimensional analog of the three-vector Maxwell
equations in the usual form, 
\begin{equation}
\nabla \times \mathbf{H}-\partial _{0}\mathbf{E}=e\mathbf{J}\qquad \qquad
\nabla \cdot \mathbf{E}=eJ^{0}  \label{432}
\end{equation}%
\begin{equation}
\nabla \cdot \mathbf{H}=0\qquad \qquad \nabla \times \mathbf{E}+\partial _{0}%
\mathbf{H}=0\;.  \label{434}
\end{equation}%
The three vector form of the field equations \cite{emlf}, defined through%
\begin{equation}
e_{i}=f^{0i}\qquad h_{i}=\epsilon _{ijk}f^{jk}  \label{440}
\end{equation}%
\begin{equation}
\epsilon ^{i}=f^{5i}\qquad \epsilon ^{0}=f^{50}  \label{450}
\end{equation}%
is useful for the study of the discrete symmetries. These field equations
(generalizations of (\ref{432}) and (\ref{434})) are%
\begin{equation}
\nabla \cdot \mathbf{e}-\partial _{\tau }\epsilon ^{0}=ej^{0}\qquad \nabla
\times \mathbf{e}+\partial _{0}\mathbf{h}=0  \label{460}
\end{equation}%
\begin{equation}
\nabla \times \mathbf{h}-\partial _{0}\mathbf{e}-\partial _{\tau }\mathbf{%
\epsilon }=e\mathbf{j}\qquad \nabla \cdot \mathbf{h}=0  \label{470}
\end{equation}%
\begin{equation}
\nabla \cdot \mathbf{\epsilon }+\partial _{0}\epsilon ^{0}=ej^{5}\qquad
\nabla \times \mathbf{\epsilon }-\sigma \partial _{\tau }\mathbf{h}=0
\label{480}
\end{equation}%
\begin{equation}
\nabla \epsilon ^{0}+\sigma \partial _{\tau }\mathbf{e}+\partial _{0}\mathbf{%
\epsilon }=0\mathbf{\;.}  \label{490}
\end{equation}

\subsubsection{Concatenation}

The connection with Maxwell theory enlarges on Stueckelberg's observation in
(\ref{160}). Under the conditions $j^{5}\rightarrow 0$ and $f^{5\mu
}\rightarrow 0$, pointwise in $x$ as $\tau \rightarrow \pm \infty $,
integration of (\ref{400}) over $\tau $, called concatenation of events into
a worldlines \cite{concat}, recovers the relations
\begin{equation}
\partial _{\nu }F^{\mu \nu }=eJ^{\mu }\qquad \qquad \epsilon ^{\mu \nu \rho
\lambda }\partial _{\mu }F_{\nu \rho }=0  \label{492}
\end{equation}%
where 
\begin{equation}
F^{\mu \nu }(x)=\int_{-\infty }^{\infty }d\tau \;f^{\mu \nu }(x,\tau )\qquad
\qquad \mathrm{and}\qquad \qquad A^{\mu }(x)=\int_{-\infty }^{\infty }d\tau
\;a^{\mu }(x,\tau )  \label{494}
\end{equation}%
and so $a^{\alpha }(x,\tau )$ has been called the pre-Maxwell field. It
follows from (\ref{494}) that $e_{0}$ and $\lambda $ have dimensions of
length.

\subsubsection{Implications for the Parameterized Mechanics}

As in the non-relativistic case, the two-body action-at-a-distance potential
in the Horwitz-Piron theory \cite{H-P} may be understood as the
approximation $-e_{0}a_{5}(x,\tau )\longrightarrow V(x)$. Within this
framework, solutions have been found for the generalizations of the standard
central force problem, including potential scattering \cite{scattering} and
bound states \cite{I,II}. Examination of radiative transitions \cite{selrul}%
, in particular the Zeeman \cite{zeeman} and Stark effects \cite{stark},
indicate that all five components of the gauge potential are necessary for
an adequate explanation of observed phenomenology.

\subsubsection{The Classical Lorentz Force}

From the quantum Hamiltonian in (\ref{170}) one is led to the classical
Lagrangian 
\begin{equation}
L=\dot{x}^{\mu }p_{\mu }-K=\frac{1}{2}M\dot{x}^{\mu }\dot{x}_{\mu }+e_{0}%
\dot{x}^{\alpha }a_{\alpha }=\frac{1}{2}M\dot{x}^{\mu }\dot{x}_{\mu }+e_{0}%
\dot{x}^{\mu }a_{\mu }+e_{0}a_{5}\;\;,  \label{500}
\end{equation}%
and, under variation with respect to $x^{\mu }$, the Lorentz force \cite%
{emlf} 
\begin{equation}
M\;\ddot{x}^{\mu }=e_{0}\ f_{\;\;\;\alpha }^{\mu }(x,\tau )\,\dot{x}^{\alpha
}=e_{0}\ \left[ f_{\;\;\;\nu }^{\mu }(x,\tau )\,\dot{x}^{\nu
}+f_{\;\;\;5}^{\mu }(x,\tau )\right] \;\;,  \label{510}
\end{equation}%
where $f_{\alpha \beta }$ is the gauge invariant quantity (\ref{370}). The
field strength $f^{5\mu }$ plays the role of Stueckelburg's field $K^{\mu }$%
, and so the enlarged gauge symmetry is seen to provide a consistent basis
for this additional interaction. The effect of this interaction on rest
mass, as Stueckelburg found earlier,%
\begin{equation}
\frac{d}{d\tau }(-\frac{1}{2}M\dot{x}^{2})=-M\dot{x}^{\mu }\ddot{x}_{\mu
}=-e_{0}\ \dot{x}^{\mu }(f_{\mu 5}+f_{\mu \nu }\dot{x}^{\nu })=-e_{0}\ \dot{x%
}^{\mu }f_{\mu 5}=e_{0}\sigma \ f_{\;\;\;\alpha }^{5}\dot{x}^{\alpha }\;\;,
\label{520}
\end{equation}%
appears formally as the \textquotedblleft fifth\textquotedblright\ component
of the Lorentz force law. Conservation of mass, $\dot{x}^{2}=constant$,
requires that%
\begin{equation}
f_{5\mu }=0\qquad \mathrm{and}\qquad \partial _{\tau }f^{\mu \nu }=0\;\;,
\label{530}
\end{equation}%
where the second condition follows from (\ref{430}) for $f^{5\mu }=0$. The
generalization of (\ref{510}) to curved spacetime was found \cite{beyond} to
be%
\begin{equation}
M[\ddot{x}^{\mu }+\Gamma ^{\mu \lambda \nu }\dot{x}_{\lambda }\dot{x}_{\nu
}]=e_{0}\ f_{\;\;\;\alpha }^{\mu }(x,\tau )\,\dot{x}^{\alpha }=e_{0}\ \left[
f_{\;\;\;\nu }^{\mu }(x,\tau )\,\dot{x}^{\nu }+f_{\;\;\;5}^{\mu }(x,\tau )%
\right]  \label{533}
\end{equation}%
with 
\begin{equation}
\Gamma ^{\mu \lambda \nu }=-{\frac{1}{2}}(\partial ^{\nu }g^{\lambda \mu
}+\partial ^{\lambda }g^{\mu \nu }-\partial ^{\mu }g^{\lambda \nu })~~.
\end{equation}%
Equation (\ref{533}) can be identified with Stueckelberg's proposed equation
(\ref{10}). This expression was shown \cite{beyond} to be the most general
expression for a classical force consistent with the quantum commutation
relations%
\begin{equation}
\left[ x^{\mu },x^{\nu }\right] =0\qquad m\left[ x^{\mu },\dot{x}^{\nu }%
\right] =-i\hbar g^{\mu \nu }\left( x\right) \;.  \label{550}
\end{equation}%
Relaxing the mass-shell constraint in (\ref{550}) breaks general
reparameterization invariance in (\ref{500}), but, under the conditions (\ref%
{530}), the remaining $\tau $-translation symmetry is associated, via
Noether's theorem, with dynamic conservation of the mass. It has been shown 
\cite{emlf} that while the material events and gauge fields may exchange
mass when the conditions (\ref{530}) do not hold, the total mass-energy of
the particles and fields is conserved. Since the gauge fields propagate with
a mass spectrum, this theory has been called off-shell electrodynamics.

\subsubsection{Classical Coulomb Problem}

Further questions of interpretation of the five dimensional formalism arise
in treating the classical Coulomb problem. Posing the classical equations of
motion for a test event (worldline $B$ in Figure 2) moving in the field
induced by a `static' event (worldline $A$ in Figure 2) evolving uniformly
along the time axis, one is faced with three interrelated problems.

\begin{center}
\includegraphics[width=250pt]{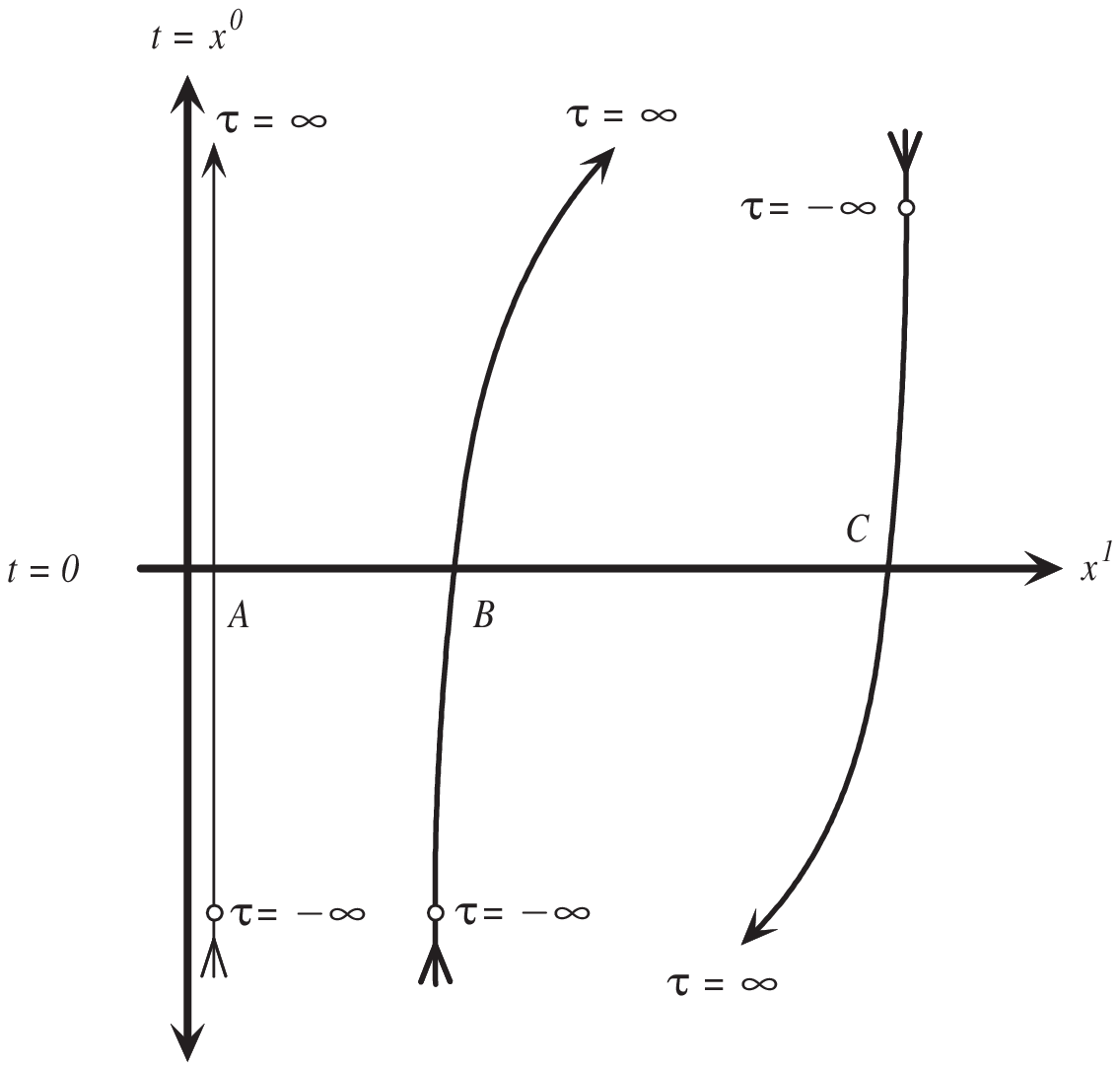}

\begin{tabular}{l}
{\footnotesize \ \textbf{Figure 2}: Elastic Scattering} \\ 
\begin{tabular}{l}
{\footnotesize A: `Static' event evolving uniformly along the time axis} \\ 
{\footnotesize B: Test event with $\dot x^0 (-\infty)=1$ evolving in field
induced by A} \\ 
{\footnotesize C: Test event with $\dot x^0 (-\infty)=-1$ evolving in field
induced by A}%
\end{tabular}%
\end{tabular}
\end{center}

First, the classical current density of the point events involves a delta
function centered on the event, 
\begin{equation}
j^{0}(x,\tau )=\delta (x^{0}-\tau )\delta ^{3}(\mathbf{x})\ \ .
\end{equation}%
Second, the structure of the Green's function \cite{green}, 
\begin{equation}
G(x,\tau )=-{\frac{1}{{4\pi }}}\delta (x^{2})\delta (\tau )-{\frac{1}{{2\pi
^{2}}}}{\frac{\partial }{{\partial {x^{2}}}}}\ {\frac{{\theta (-\sigma
g_{\alpha \beta }x^{\alpha }x^{\beta })}}{\sqrt{-\sigma g_{\alpha \beta
}x^{\alpha }x^{\beta }}}}  \label{560}
\end{equation}%
for the five dimensional wave equation \cite{saad} 
\begin{equation}
\partial _{\alpha }\partial ^{\alpha }f^{\beta \gamma }=(\partial _{\mu
}\partial ^{\mu }+\partial _{\tau }\partial ^{\tau })f^{\beta \gamma
}=(\partial _{\mu }\partial ^{\mu }+\sigma \;\partial _{\tau }^{2})f^{\beta
\gamma }=-e(\partial ^{\beta }j^{\gamma }-\partial ^{\gamma }j^{\beta })
\label{570}
\end{equation}%
carries these delta functions into the induced gauge potentials. Third, the $%
\tau $-translation symmetry of the asymptotic events leads to a strong
dependence on initial conditions that, under concatenation, should not be
observable; that is, scattering in the gauge field induced by the event $%
x^{\mu }\left( \tau -a\right) $ will depend qualitatively on the value of $a$%
, even though the concatenated Coulomb field does not. A reasonable approach
was found \cite{larry} by smoothing the field-inducing current as 
\begin{equation}
j_{\varphi }^{\alpha }(x,\tau )=\int_{-\infty }^{\infty }ds\ \varphi (\tau
-s)\ j^{\alpha }(x,s)  \label{580}
\end{equation}%
where $\varphi (\tau )$ is the Laplace distribution 
\begin{equation}
\varphi (\tau )=\frac{1}{2\lambda }e^{-|\tau |/\lambda }\;\;.  \label{590}
\end{equation}%
Since the smoothing distribution satisfies 
\begin{equation}
\int_{-\infty }^{\infty }d\tau \ \varphi (\tau )=1  \label{600}
\end{equation}%
the concatenated Maxwell current is not affected by the integration (\ref%
{580}). This approach leads, in the low energy case, to a classical Yukawa
potential, 
\begin{equation}
a_{\varphi }^{0}(x,\tau )=\frac{1}{2\lambda }\left[ -\frac{e}{4\pi R}\
e^{-R/\lambda }\right] \left[ \frac{1}{2}\left( {{\frac{dx^{0}}{d\tau }}}%
+1\right) \right] \simeq \frac{1}{2\lambda }\left[ -\frac{e}{4\pi R}\
e^{-R/\lambda }\right] ,  \label{610}
\end{equation}%
\begin{equation}
M\;{{\frac{d^{2}\mathbf{x}}{d\tau ^{2}}}}\simeq 2\lambda ~e\ \mathbf{\nabla ~%
}a_{\varphi }^{0}(x,\tau )=\mathbf{\nabla }\left[ -\frac{e^{2}}{4\pi R}\
e^{-R/\lambda }\right] \mathbf{~}\quad ,  \label{620}
\end{equation}%
with interpretation of $\lambda $ as a cut-off in the mass spectrum of the
photons mediating the interaction. It was shown in \cite{high-order} that
replacing the usual kinetic term for the electromagnetic field strengths in (%
\ref{360})%
\begin{equation}
S_{em-kinetic}^{0}=\frac{\lambda }{4}\int d^{4}x\,d\tau \,f^{\alpha \beta
}(x,\tau )\,\ f_{\alpha \beta }(x,\tau )\,  \label{630}
\end{equation}%
with the higher-order derivative term 
\begin{equation}
S_{em-kinetic}=S_{em-kinetic}^{0}+\frac{\lambda ^{3}}{4}\int d^{4}x\,d\tau
\,\ 
\Bigl%
(\partial _{\tau }f^{\alpha \beta }(x,\tau )%
\Bigr%
)%
\Bigl%
(\partial _{\tau }\ f_{\alpha \beta }\left( x,\tau \right) \,%
\Bigr%
)\,  \label{640}
\end{equation}%
is equivalent to the \emph{ad hoc} approach taken in (\ref{580}). This
approach puts the theory into a form amenable to quantization and provides
the mass cut-off which makes the field theory finite at all orders of
perturbation theory.

\subsection{Particle/Antiparticle Problem}

By Stueckelberg's interpretation of the worldlines, application of the
classical Coulomb case to elastic particle-antiparticle scattering seems
only to require that the initial conditions of the test event include $%
dx^{0}/d\tau \leq 0$ (worldline $C$ in Figure 2 shows an event evolving from
initial condition $t=-\tau$). However, in two aspects, the solution to the
scattering problem posed in this way seems inadequate. First, we notice that
in the low energy case, $dx^{0}/d\tau \simeq -1$, the Lorentz force (\ref%
{610}) takes the form%
\begin{equation}
a_{\varphi }^{0}(x,\tau )=\frac{1}{2\lambda }\left[ -\frac{e}{4\pi R}\
e^{-R/\lambda }\right] \left[ \frac{1}{2}\left( {{\frac{dx^{0}}{d\tau }}}%
+1\right) \right] \simeq 0\;,  \label{660}
\end{equation}%
and it appears that no scattering takes place. Second, in modern treatments
(for example \cite{novozhilov, IZ}), based on the $CPT$ theorem, the
antiparticle is characterized as a particle with the signs of all its
additive quantum numbers reversed. Following Wigner, the quantum reflection $%
\theta $, which both reverses the spacetime parameters and exchanges the
past and future, is an antiunitary operator. Since $\theta $ must be
independent of any internal symmetries associated with generators $\{F_{n}\}$%
, the antiunitary of $\theta $ requires that%
\begin{equation}
\theta e^{iF_{n}\alpha ^{n}}\theta ^{-1}=e^{iF_{n}\alpha ^{n}}\quad
\Rightarrow \quad \theta \left[ iF_{\mu }\right] \theta ^{-1}=-i\left[
\theta F_{\mu }\theta ^{-1}\right] \quad \Rightarrow \quad \theta F_{\mu
}\theta ^{-1}=-F_{\mu }  \label{670}
\end{equation}%
and so all additive quantum numbers change sign for the antiparticle.
However, in the Stueckelberg theory, states are labeled by the parameter $%
\tau $, which is not affected by spacetime inversion, and the argument is
not immediately applicable to the event evolving toward earlier $t$. It is
not clear how this general reflection of internal symmetries is related to
the dynamic evolution of the energy. For these reasons, we are led to
reconsider whether the antiparticle differs from the particle only in the
sign of $dx^{0}/d\tau $. In this paper, we address these issues by examining
the discrete symmetries in a formal manner.

\section{Discrete Symmetries of Off-Shell Electrodynamics}

We take the improper Lorentz transformations, in the passive sense, as
reversing the orientation of the space and time axes. This approach is
similar to the standard approach, but differs from the non-relativistic
formulation following Wigner \cite{movies} (see also \cite{Jackson}), in
which time reversal changes the sign of velocities, momenta, and the space
part of currents. Since velocities are $\tau$-derivatives, they transform
tensorially under the improper Lorentz transformations, and so time and
space inversion are handled as coordinates on the same footing. Reversal of
the sense of motion, which is part of time reversal in the standard
treatment, would be associated with $\tau$-reversal in this theory. However,
since $\tau$-reversal cannot be implemented as a Lorentz transformation, we
do not assume that the theory must be form covariant under such an operation.

Then space inversion acts as 
\begin{equation}
x=\left( x^{0},\mathbf{x}\right) \mathrel{\mathop{\longrightarrow
}\limits_{P}}x_{P}=\left( x_{P}^{0},\mathbf{x}_{P}\right) =\left( x^{0},-%
\mathbf{x}\right)
\end{equation}%
and time inversion acts as 
\begin{equation}
x=\left( x^{0},\mathbf{x}\right) \mathrel{\mathop{\longrightarrow
}\limits_{T}}x_{T}=\left( x_{T}^{0},\mathbf{x}_{T}\right) =\left( -x^{0},%
\mathbf{x}\right) \,\,\,\,.
\end{equation}%
From the conventional observation that electromagnetic interactions are
independent of the orientation of reference frames, we expect that the
pre-Maxwell equations expressed in the coordinates $x_{P}$ or $x_{T}$ will
be identical in form to the equations expressed in $x$ coordinates. In order
to clarify our method, we take a familiar example from Maxwell theory, where
form invariance of the Coulomb law implies that 
\begin{equation}
\nabla \cdot \mathbf{E}\left( t,\mathbf{x}\right) =\rho \left( t,\mathbf{x}%
\right) \mathrel{\mathop{\longrightarrow }\limits_{P}}\nabla _{P}\cdot 
\mathbf{E}_{P}\left( t_{P},\mathbf{x}_{P}\right) =\rho _{P}\left( t_{P},%
\mathbf{x}_{P}\right) \quad .  \label{max_coul}
\end{equation}%
Since $\rho \left( t,\mathbf{x}\right) $ is a scalar field on spacetime, we
expect that its value at the point $\left( t,\mathbf{x}\right) $ equals the
value of the transformed field $\rho _{P}$ at the corresponding point $%
\left( t_{P},\mathbf{x}_{P}\right) $, that is 
\begin{equation}
\rho _{P}\left( t_{P},\mathbf{x}_{P}\right) =\rho \left( t,\mathbf{x}\right)
\quad .
\end{equation}%
Therefore, 
\begin{eqnarray}
\nabla _{P}\cdot \mathbf{E}_{P}\left( t_{P},\mathbf{x}_{P}\right) &=&\rho
\left( t,\mathbf{x}\right) \\
-\nabla \cdot \mathbf{E}_{P}\left( t,-\mathbf{x}\right) &=&\rho \left( t,%
\mathbf{x}\right) \\
\nabla \cdot \left\{ -\mathbf{E}_{P}\left( t,-\mathbf{x}\right) \right\}
&=&\rho \left( t,\mathbf{x}\right) \quad .  \label{P-coul}
\end{eqnarray}%
Comparing (\ref{P-coul}) with (\ref{max_coul}) we notice that $\mathbf{E}%
\left( t,\mathbf{x}\right) $ and $-\mathbf{E}_{P}\left( t,-\mathbf{x}\right) 
$ satisfy the same equation, so conclude 
\begin{equation}
\mathbf{E}\left( t,\mathbf{x}\right) =-\mathbf{E}_{P}\left( t,-\mathbf{x}%
\right) =-\mathbf{E}_{P}\left( t_{P},\mathbf{x}_{P}\right) \quad .
\end{equation}%
Standard treatments of the space and time reversal properties of the Maxwell
theory (see for example \cite{Jackson}) study the field equations under the
assumption of orientation invariance. However, since pre-Maxwell
electrodynamics emerged from the requirement of local gauge invariance, from
which the classical Lorentz force follows directly, it is more appropriate
to begin with (\ref{510}).

\subsection{Space Inversion}

Under $P$, the pre-Maxwell Lorentz equations in three-vector form 
\begin{eqnarray}
M\frac{d^{2}x^{0}}{d\tau ^{2}} &=&e_{0}\left[ \mathbf{e}\left( t,\mathbf{x}%
,\tau \right) \cdot \frac{d\mathbf{x}}{d\tau }-\sigma \epsilon ^{0}\left( t,%
\mathbf{x},\tau \right) \right]  \label{lor0-1} \\
M\frac{d^{2}\mathbf{x}}{d\tau ^{2}} &=&e_{0}\left[ \mathbf{e}\left( t,%
\mathbf{x},\tau \right) \,\,\frac{dx^{0}}{d\tau }+\frac{d\mathbf{x}}{d\tau }%
\times \mathbf{h}\left( t,\mathbf{x},\tau \right) -\sigma \mathbf{\epsilon }%
\left( t,\mathbf{x},\tau \right) \right]  \label{lor2-1}
\end{eqnarray}%
become 
\begin{eqnarray}
M\frac{d^{2}x_{P}^{0}}{d\tau ^{2}} &=&e_{0}\left[ \mathbf{e}_{P}\left( t_{P},%
\mathbf{x}_{P},\tau \right) \cdot \frac{d\mathbf{x}_{P}}{d\tau }-\sigma
\epsilon _{P}^{0}\left( t_{P},\mathbf{x}_{P},\tau \right) \right] \\
M\frac{d^{2}\mathbf{x}_{P}}{d\tau ^{2}} &=&e_{0}\left[ \mathbf{e}_{P}\left(
t_{P},\mathbf{x}_{P},\tau \right) \frac{dx_{P}^{0}}{d\tau }+\frac{d\mathbf{x}%
_{P}}{d\tau }\times \mathbf{h}_{P}\left( t_{P},\mathbf{x}_{P},\tau \right)
-\sigma \mathbf{\epsilon }_{P}\left( t_{P},\mathbf{x}_{P},\tau \right) %
\right]
\end{eqnarray}%
so that 
\begin{eqnarray}
M\frac{d^{2}x^{0}}{d\tau ^{2}} &=&e_{0}\left[ \mathbf{e}_{P}\left( t_{P},%
\mathbf{x}_{P},\tau \right) \cdot \left( -\frac{d\mathbf{x}}{d\tau }\right)
-\sigma \epsilon _{P}^{0}\left( t_{P},\mathbf{x}_{P},\tau \right) \right] \\
-M\frac{d^{2}\mathbf{x}}{d\tau ^{2}} &=&e_{0}\left[ \mathbf{e}_{P}\left(
t_{P},\mathbf{x}_{P},\tau \right) \frac{dx^{0}}{d\tau }-\frac{d\mathbf{x}}{%
d\tau }\times \mathbf{h}_{P}\left( t_{P},\mathbf{x}_{P},\tau \right) -\sigma 
\mathbf{\epsilon }_{P}\left( t_{P},\mathbf{x}_{P},\tau \right) \right]
\end{eqnarray}%
and finally 
\begin{eqnarray}
M\frac{d^{2}x^{0}}{d\tau ^{2}} &=&e_{0}\left[ \left( -\mathbf{e}_{P}\left(
t_{P},\mathbf{x}_{P},\tau \right) \right) \cdot \frac{d\mathbf{x}}{d\tau }%
-\sigma \left( \epsilon _{P}^{0}\left( t_{P},\mathbf{x}_{P},\tau \right)
\right) \right]  \label{P-lor0} \\
M\frac{d^{2}\mathbf{x}}{d\tau ^{2}} &=&e_{0}\left[ \left( -\mathbf{e}%
_{P}\left( t_{P},\mathbf{x}_{P},\tau \right) \right) \frac{dx^{0}}{d\tau }+%
\frac{d\mathbf{x}}{d\tau }\times \left( \mathbf{h}_{P}\left( t_{P},\mathbf{x}%
_{P},\tau \right) \right) -\sigma \left( -\mathbf{\epsilon }\left( t_{P},%
\mathbf{x}_{P},\tau \right) \right) \right] \quad \quad .  \label{P-lor2}
\end{eqnarray}%
Comparing (\ref{lor0-1}) and (\ref{lor2-1}) with (\ref{P-lor0}) and (\ref%
{P-lor2}) shows that 
\begin{eqnarray}
\mathbf{e}_{P}\left( t_{P},\mathbf{x}_{P},\tau \right) &=&-\mathbf{e}\left(
t,\mathbf{x},\tau \right)  \label{e-p} \\
\mathbf{h}_{P}\left( t_{P},\mathbf{x}_{P},\tau \right) &=&\mathbf{h}\left( t,%
\mathbf{x},\tau \right)  \label{h-p} \\
\epsilon _{P}^{0}\left( t_{P},\mathbf{x}_{P},\tau \right) &=&\epsilon
^{0}\left( t,\mathbf{x},\tau \right)  \label{eps-p} \\
\mathbf{\epsilon }_{P}\left( t_{P},\mathbf{x}_{P},\tau \right) &=&-\mathbf{%
\epsilon }\left( t,\mathbf{x},\tau \right)  \label{eps-p0}
\end{eqnarray}%
and we recognize (\ref{e-p}) and (\ref{h-p}) as the behavior of the Maxwell
electric and magnetic 3-vectors under parity.

\subsection{Time Inversion}

Under $T$, (\ref{lor0-1}) and (\ref{lor2-1}) become 
\begin{eqnarray}
M\frac{d^{2}x_{T}^{0}}{d\tau ^{2}} &=&e_{0}\left[ \mathbf{e}_{T}\left( t_{T},%
\mathbf{x}_{T},\tau \right) \cdot \frac{d\mathbf{x}_{T}}{d\tau }-\sigma
\epsilon _{T}^{0}\left( t_{T},\mathbf{x}_{T},\tau \right) \right] \\
M\frac{d^{2}\mathbf{x}_{T}}{d\tau ^{2}} &=&e_{0}\left[ \mathbf{e}_{T}\left(
t_{T},\mathbf{x}_{T},\tau \right) \frac{dx_{T}^{0}}{d\tau }+\frac{d\mathbf{x}%
_{T}}{d\tau }\times \mathbf{h}_{T}\left( t_{T},\mathbf{x}_{T},\tau \right)
-\sigma \mathbf{\epsilon }_{T}\left( t_{T},\mathbf{x}_{T},\tau \right) %
\right]
\end{eqnarray}%
so that 
\begin{eqnarray}
-M\frac{d^{2}x^{0}}{d\tau ^{2}} &=&e_{0}\left[ \mathbf{e}_{T}\left( t_{T},%
\mathbf{x}_{T},\tau \right) \cdot \frac{d\mathbf{x}}{d\tau }-\sigma \epsilon
_{T}^{0}\left( t_{T},\mathbf{x}_{T},\tau \right) \right] \\
M\frac{d^{2}\mathbf{x}}{d\tau ^{2}} &=&e_{0}\left[ \mathbf{e}_{T}\left(
t_{T},\mathbf{x}_{T},\tau \right) \left( -\frac{dx^{0}}{d\tau }\right) +%
\frac{d\mathbf{x}}{d\tau }\times \mathbf{h}_{T}\left( t_{T},\mathbf{x}%
_{T},\tau \right) -\sigma \mathbf{\epsilon }_{T}\left( t_{T},\mathbf{x}%
_{T},\tau \right) \right]
\end{eqnarray}%
and finally 
\begin{eqnarray}
M\frac{d^{2}x^{0}}{d\tau ^{2}} &=&e_{0}\left[ \left( -\mathbf{e}_{T}\left(
t_{T},\mathbf{x}_{T},\tau \right) \right) \cdot \frac{d\mathbf{x}}{d\tau }%
-\sigma \left( -\epsilon _{T}^{0}\left( t_{T},\mathbf{x}_{T},\tau \right)
\right) \right] \\
M\frac{d^{2}\mathbf{x}}{d\tau ^{2}} &=&e_{0}\left[ \left( -\mathbf{e}%
_{T}\left( t_{T},\mathbf{x}_{T},\tau \right) \right) \frac{dx^{0}}{d\tau }+%
\frac{d\mathbf{x}}{d\tau }\times \left( \mathbf{h}_{T}\left( t_{T},\mathbf{x}%
_{T},\tau \right) \right) -\sigma \left( \mathbf{\epsilon }_{T}\left( t_{T},%
\mathbf{x}_{T},\tau \right) \right) \right] \quad \quad .
\end{eqnarray}%
Comparing (\ref{lor0-1}) and (\ref{lor2-1}) with (\ref{P-lor0}) and (\ref%
{P-lor2}) shows that 
\begin{eqnarray}
\mathbf{e}_{T}\left( t_{T},\mathbf{x}_{T},\tau \right) &=&-\mathbf{e}\left(
t,\mathbf{x},\tau \right)  \label{e-t} \\
\mathbf{h}_{T}\left( t_{T},\mathbf{x}_{T},\tau \right) &=&\mathbf{h}\left( t,%
\mathbf{x},\tau \right)  \label{h-t} \\
\epsilon _{T}^{0}\left( t_{T},\mathbf{x}_{T},\tau \right) &=&-\epsilon
^{0}\left( t,\mathbf{x},\tau \right)  \label{eps-0t} \\
\mathbf{\epsilon }_{T}\left( t_{T},\mathbf{x}_{T},\tau \right) &=&\mathbf{%
\epsilon }\left( t,\mathbf{x},\tau \right)  \label{eps-t}
\end{eqnarray}%
and here we notice that (\ref{e-t}) and (\ref{h-t}) are \emph{opposite} to
the standard behavior of the electric and magnetic 3-vectors under time
inversion. This opposite behavior can be attributed to our having respected
the independence of $x^{0}\left( \tau \right) $ as a function of $\tau $,
not constrained by the mass-shell condition 
\begin{equation}
\frac{dx^{0}}{d\tau }=+\frac{1}{\sqrt{1-\left( d\mathbf{x}/dt\right) ^{2}}}%
\,\,\,\,\,.
\end{equation}

\subsection{Currents}

The pre-Maxwell equations in 3-vector form are 
\begin{eqnarray}
\nabla \cdot \mathbf{e}-\partial _{\tau }\epsilon ^{0} &=&ej^{0}
\label{inhom1} \\
\nabla \times \mathbf{h}-\partial _{0}\mathbf{e}-\partial _{\tau }\mathbf{%
\epsilon } &=&e\mathbf{j}  \label{inhom2} \\
\nabla \cdot \mathbf{\epsilon }+\partial _{0}\epsilon ^{0} &=&ej^{5}
\label{inhom3} \\
\nabla \cdot \mathbf{h} &=&0  \label{hom1} \\
\nabla \times \mathbf{e}+\partial _{0}\mathbf{h} &=&0  \label{hom2} \\
\nabla \times \mathbf{\epsilon }-\sigma \partial _{\tau }\mathbf{h} &=&0
\label{hom3} \\
\nabla \epsilon ^{0}+\sigma \partial _{\tau }\mathbf{e}+\partial _{0}\mathbf{%
\ \epsilon } &=&0  \label{hom4}
\end{eqnarray}%
so that under space inversion $P$, they become 
\begin{eqnarray}
\left( -\nabla \right) \cdot \left( -\mathbf{e}\right) -\partial _{\tau
}\epsilon ^{0} &=&\nabla \cdot \mathbf{e}-\partial _{\tau }\epsilon
^{0}=ej_{P}^{0} \\
\left( -\nabla \right) \times \mathbf{h}-\partial _{0}\left( -\mathbf{e}%
\right) -\partial _{\tau }\left( -\mathbf{\epsilon }\right) &=&-\left[
\nabla \times \mathbf{h}-\partial _{0}\mathbf{e}-\partial _{\tau }\mathbf{%
\epsilon }\right] =e\mathbf{j}_{P} \\
\left( -\nabla \right) \cdot \left( -\mathbf{\epsilon }\right) +\partial
_{0}\epsilon ^{0} &=&\nabla \cdot \mathbf{\epsilon }+\partial _{0}\epsilon
^{0}=ej_{P}^{5} \\
\left( -\nabla \right) \cdot \mathbf{h} &=&-\left[ \nabla \cdot \mathbf{h}%
\right] =0 \\
\left( -\nabla \right) \times \left( -\mathbf{e}\right) +\partial _{0}%
\mathbf{h} &=&\nabla \times \mathbf{e}+\partial _{0}\mathbf{h}=0 \\
\left( -\nabla \right) \times \left( -\mathbf{\epsilon }\right) -\sigma
\partial _{\tau }\mathbf{h} &=&\nabla \times \mathbf{\epsilon }-\sigma
\partial _{\tau }\mathbf{\ h}=0 \\
\left( -\nabla \right) \epsilon ^{0}+\sigma \partial _{\tau }\left( -\mathbf{%
e}\right) +\partial _{0}\left( -\mathbf{\epsilon }\right) &=&-\left[ \nabla
\epsilon ^{0}+\sigma \partial _{\tau }\mathbf{e}+\partial _{0}\mathbf{%
\epsilon }\right] =0
\end{eqnarray}%
which are form invariant under the choices 
\begin{eqnarray}
j_{P}^{0}\left( t_{P},\mathbf{x}_{P},\tau \right) &=&j^{0}\left( t,\mathbf{x}%
,\tau \right) \\
\mathbf{j}_{P}\left( t_{P},\mathbf{x}_{P},\tau \right) &=&-\mathbf{j}\left(
t,\mathbf{x},\tau \right) \\
j_{P}^{5}\left( t_{P},\mathbf{x}_{P},\tau \right) &=&j^{5}\left( t,\mathbf{x}%
,\tau \right) \,\,\,\,\,\,\,\,\,.
\end{eqnarray}%
Similarly, under $T$, 
\begin{eqnarray}
\nabla \cdot \left( -\mathbf{e}\right) -\partial _{\tau }\left( -\epsilon
^{0}\right) &=&-\left[ \nabla \cdot \mathbf{e}-\partial _{\tau }\epsilon ^{0}%
\right] =ej_{T}^{0} \\
\nabla \times \mathbf{h}-\left( -\partial _{0}\right) \left( -\mathbf{e}%
\right) -\partial _{\tau }\mathbf{\epsilon } &=&\nabla \times \mathbf{h}%
-\partial _{0}\mathbf{\ e}-\partial _{\tau }\mathbf{\epsilon }=e\mathbf{j}%
_{T} \\
\nabla \cdot \mathbf{\epsilon }+\left( -\partial _{0}\right) \left(
-\epsilon ^{0}\right) &=&\nabla \cdot \mathbf{\epsilon }+\partial
_{0}\epsilon ^{0}=ej_{T}^{5} \\
\nabla \cdot \mathbf{h} &=&0 \\
\nabla \times \left( -\mathbf{e}\right) +\left( -\partial _{0}\right) 
\mathbf{h} &=&-\left[ \nabla \times \mathbf{e}+\partial _{0}\mathbf{h}\right]
=0 \\
\nabla \times \mathbf{\epsilon }-\sigma \partial _{\tau }\mathbf{h} &=&0 \\
\nabla \left( -\epsilon ^{0}\right) +\sigma \partial _{\tau }\left( -\mathbf{%
e}\right) +\left( -\partial _{0}\right) \mathbf{\epsilon } &=&-\left[ \nabla
\epsilon ^{0}+\sigma \partial _{\tau }\mathbf{e}+\partial _{0}\mathbf{%
\epsilon }\right] =0
\end{eqnarray}%
which are form invariant under the choices 
\begin{eqnarray}
j_{T}^{0}\left( t_{T},\mathbf{x}_{T},\tau \right) &=&-j^{0}\left( t,\mathbf{x%
},\tau \right) \\
\mathbf{j}_{T}\left( t_{T},\mathbf{x}_{T},\tau \right) &=&\mathbf{j}\left( t,%
\mathbf{x},\tau \right) \\
j_{T}^{5}\left( t_{T},\mathbf{x}_{T},\tau \right) &=&j^{5}\left( t,\mathbf{x}%
,\tau \right) \,\,\,\,\,\,\,\,\,.
\end{eqnarray}%
From the transformation properties for the field strengths, we may deduce
the transformation properties of the 5-vector potential components. From 
\begin{equation}
f^{\alpha \beta }=\partial ^{\alpha }a^{\beta }-\partial ^{\beta }a^{\alpha }
\end{equation}%
we have 
\begin{equation}
\mathbf{e}^{i}=\partial ^{0}a^{i}-\partial ^{i}a^{0}\mathrel{\mathop{%
\rightarrow }\limits_{P}}-\mathbf{e}^{i}=\partial ^{0}a_{P}^{i}-\left(
-\partial ^{i}\right) a_{P}^{0}=-\left( \partial ^{0}a^{i}-\partial
^{i}a^{0}\right)
\end{equation}%
so 
\begin{equation}
a_{P}^{0}=a^{0}\,\,\,\,\,\,\,\,\,\,\,\,\,\,\,\,\,\,\,\,\,\,\,%
\,a_{P}^{i}=-a^{i}
\end{equation}%
consistent with 
\begin{equation}
\mathbf{h}^{i}=\varepsilon ^{ijk}\partial _{j}a_{k}\;.  \label{h-def}
\end{equation}%
Similarly, 
\begin{equation}
\mathbf{e}^{i}=\partial ^{0}a^{i}-\partial ^{i}a^{0}\mathrel{\mathop{%
\rightarrow }\limits_{T}}-\mathbf{e}^{i}=\left( -\partial ^{0}\right)
a_{T}^{i}-\partial ^{i}a_{T}^{0}=-\left( \partial ^{0}a^{i}-\partial
^{i}a^{0}\right)
\end{equation}%
so 
\begin{equation}
a_{T}^{0}=-a^{0}\,\,\,\,\,\,\,\,\,\,\,\,\,\,\,\,\,\,\,\,\,\,\,%
\,a_{T}^{i}=a^{i}
\end{equation}%
again consistent with (\ref{h-def}). For the second vector field, 
\begin{equation}
\mathbf{\epsilon }^{i}=\partial ^{5}a^{i}-\partial ^{i}a^{5}%
\mathrel{\mathop{ \rightarrow }\limits_{P}}-\mathbf{\epsilon }^{i}=\partial
^{5}a_{P}^{i}-\left( -\partial ^{i}\right) a_{P}^{5}=-\left( \partial
^{5}a^{i}-\partial ^{i}a^{5}\right)
\end{equation}%
leads to 
\begin{equation}
a_{P}^{5}=a^{5}\,\,\,\,\,\,\,\,\,\,\,\,\,\,\,\,\,\,\,\,\,\,\,\,\,\,%
\,a_{P}^{i}=-a^{i}
\end{equation}%
which is consistent with 
\begin{equation}
\epsilon ^{0}=\partial ^{5}a^{0}-\partial ^{0}a^{5}  \label{eps-0}
\end{equation}%
and 
\begin{equation}
\mathbf{\epsilon }^{i}=\partial ^{5}a^{i}-\partial ^{i}a^{5}\,\,.
\label{eps-i}
\end{equation}%
Similarly 
\begin{equation}
\mathbf{\epsilon }^{i}=\partial ^{5}a^{i}-\partial ^{i}a^{5}%
\mathrel{\mathop{ \rightarrow }\limits_{T}}\mathbf{\epsilon }^{i}=\partial
^{5}a_{T}^{i}-\left( \partial ^{i}\right) a_{T}^{5}=\left( \partial
^{5}a^{i}-\partial ^{i}a^{5}\right)
\end{equation}%
requires 
\begin{equation}
a_{T}^{5}=a^{5}\,\,\,\,\,\,\,\,\,\,\,\,\,\,\,\,\,\,\,\,\,\,\,\,\,\,%
\,a_{T}^{i}=a^{i}\quad .
\end{equation}%
All of the 5-vector quantities encountered up to this point transform
tensorially under space and time inversion, as the quantity $\left( x^{0},%
\mathbf{x},\tau \right) $. We summarize the results thus far in Table 1,

\begin{center}
\begin{tabular}[t]{|l|c|c|}
\hline
\quad \rule[-0.4cm]{0cm}{1cm}\textbf{Quantity\quad } & Transformation Under $%
P$ & Transformation Under $T$ \\ \hline
$\left( x^{0},\mathbf{x},\tau \right) \rule[-0.4cm]{0cm}{1cm}$ & $\left(
x^{0},-\mathbf{x},\tau \right) $ & $\left( -x^{0},\mathbf{x},\tau \right) $
\\ \hline
\rule[-0.4cm]{0cm}{1cm}$\mathbf{e}$ & $-\mathbf{e}$ & $-\mathbf{e}$ \\ \hline
\rule[-0.4cm]{0cm}{1cm}$\mathbf{h}$ & $\mathbf{h}$ & $\mathbf{h}$ \\ \hline
\rule[-0.4cm]{0cm}{1cm}$\mathbf{\epsilon }$ & $-\mathbf{\epsilon }$ & $%
\mathbf{\epsilon }$ \\ \hline
\rule[-0.4cm]{0cm}{1cm}$\epsilon ^{0}$ & $\epsilon ^{0}$ & $-\epsilon ^{0}$
\\ \hline
\rule[-0.4cm]{0cm}{1cm}$\left( j^{0},\mathbf{j},j^{5}\right) $ & $\left(
j^{0},-\mathbf{j},j^{5}\right) $ & $\left( -j^{0},\mathbf{j},j^{5}\right) $
\\ \hline
\rule[-0.4cm]{0cm}{1cm}$\left( a^{0},\mathbf{a},a^{5}\right) $ & $\left(
a^{0},-\mathbf{a},a^{5}\right) $ & $\left( -a^{0},\mathbf{a},a^{5}\right) $
\\ \hline
\end{tabular}

Table 1
\end{center}

\section{Off-Shell Quantum Mechanics}

We now turn to the discrete symmetries of the Schrodinger equation 
\begin{eqnarray}
(i\partial _{\tau }+e_{0}a_{5})\ \psi (x,\tau ) &=&\frac{1}{2M}(p^{\mu
}-e_{0}a^{\mu })(p_{\mu }-e_{0}a_{\mu })\psi (x,\tau )\ \  \\
&=&-\frac{1}{2M}(\partial ^{\mu }-ie_{0}a^{\mu })(\partial _{\mu
}-ie_{0}a_{\mu })\ \psi (x,\tau )\,\,\,\,.  \label{schrodinger}
\end{eqnarray}%
In the space reversed coordinates, the transformed equation satisfies 
\begin{equation}
(i\partial _{\tau }+e_{0}a_{5P})\ \psi _{P}(x_{P},\tau )=-\frac{1}{2M}%
(\partial _{P}^{\mu }-ie_{0}a_{P}^{\mu })(\partial _{\mu P}-ie_{0}a_{\mu
P})\ \psi _{P}(x_{P},\tau )\,\,\,\,.
\end{equation}%
From Table 1 --- and the fact that $\left( \partial _{0},\partial
_{k},\partial _{\tau }\right) $ transforms as $\left(
x_{0},x_{k},x_{5}\right) $ --- we have 
\begin{eqnarray}
(i\partial _{\tau }+e_{0}a_{5})\psi ^{P}(x_{P},\tau ) &=&  \nonumber \\
&=&-\frac{1}{2M}\Bigl[(\partial _{P}^{k}-ie_{0}a_{P}^{k})(\partial
_{kP}-ie_{0}a_{kP}) \\
&&%
\mbox{\qquad \ \ \ \ }%
-(\partial _{P}^{0}-ie_{0}a_{P}^{0})(\partial _{0P}-ie_{0}a_{0P})\Bigr]\
\psi ^{P}(x_{P},\tau )\,\,\,\, \\
&=&-\frac{1}{2M}\Bigl[(-\partial ^{k}+ie_{0}a^{k})(-\partial
_{k}+ie_{0}a_{k})  \nonumber \\
&&%
\mbox{\qquad \ \ \ \ }%
-(\partial ^{0}-ie_{0}a^{0})(\partial _{0}-ie_{0}a_{0})\Bigr ]\ \psi
^{P}(x_{P},\tau )\, \\
&=&-\frac{1}{2M}(\partial ^{\mu }-ie_{0}a^{\mu })(\partial _{\mu
}-ie_{0}a_{\mu })\ \psi ^{P}(x_{P},\tau )  \label{sh_p}
\end{eqnarray}%
Since equation (\ref{sh_p}) is explicitly identical in form to (\ref%
{schrodinger}), the solutions must be identical, so 
\begin{equation}
\psi _{P}(x_{P},\tau )=\psi (x,\tau )\quad \Rightarrow \quad \psi
_{P}(x,\tau )=\psi (x^{0},-\mathbf{x},\tau )\quad .
\end{equation}%
The corresponding argument for time inversion leads to 
\begin{equation}
\psi _{T}(x_{T},\tau )=\psi (x,\tau )\quad \Rightarrow \quad \psi
_{T}(x,\tau )=\psi (-x^{0},\mathbf{x},\tau )\quad ,
\end{equation}%
and we see that the manifest invariance under space and time inversion found
for the Horwitz-Piron theory with standard Maxwell fields \cite{concat}
applies in the presence of pre-Maxwell fields. Seeking a solution for the
replacement $e_{0}\rightarrow -e_{0}$, that is 
\begin{equation}
(i\partial _{\tau }-e_{0}a_{5})\ \psi _{C}(x,\tau )=-\frac{1}{2M}(\partial
^{\mu }+ie_{0}a^{\mu })(\partial _{\mu }+ie_{0}a_{\mu })\ \psi _{C}(x,\tau
)\,\,\,\,,  \label{sh_c}
\end{equation}%
the usual strategy for Schrodinger-like equations with minimal coupling to a
gauge field begins with complex conjugation, which expresses the antiunitary
character of the total reflection discussed in the previous chapter. In the
present case, complex conjugation of (\ref{schrodinger}) leads to 
\begin{equation}
(-i\partial _{\tau }-e_{0}a_{5})\ \psi ^{\ast }(x,\tau )=-\frac{1}{2M}%
(\partial ^{\mu }+ie_{0}a^{\mu })(\partial _{\mu }+ie_{0}a_{\mu })\ \psi
^{\ast }(x,\tau )\,\,\,\,,  \label{star-1}
\end{equation}%
which is not yet in the form (\ref{sh_c}). Unlike the case of Maxwell
fields, the possible dependence of the pre-Maxwell fields on $\tau $
prevents us from simply taking $\tau \rightarrow -\tau $ in order to reverse
the sign of $-i\partial _{\tau }$. Instead, we posit the existence of a $%
\tau $-inversion operation $\mathcal{T}$ and investigate the requirements
which make it reasonable.

Applying $\tau $-inversion to (\ref{star-1}), we find 
\begin{equation}
(i\partial _{\tau }-e_{0}a_{5\mathcal{T}})\ \psi ^{\ast }(x,-\tau )=-\frac{1%
}{2M}(\partial ^{\mu }+ie_{0}a_{\mathcal{T}}^{\mu })(\partial _{\mu
}+ie_{0}a_{\mu \mathcal{T}})\ \psi ^{\ast }(x,-\tau )\,\,\,\,,
\label{star-2}
\end{equation}%
which will be in the form of (\ref{sh_c}) if 
\begin{eqnarray}
a_{\mathcal{T}}^{\mu }(x_{\mathcal{T}},\tau _{\mathcal{T}}) &=&a^{\mu
}(x,\tau )\quad \Rightarrow \quad a_{\mathcal{T}}^{\mu }(x,\tau )=a^{\mu
}(x,-\tau ) \\
a_{5\mathcal{T}}(x_{\mathcal{T}},\tau _{\mathcal{T}}) &=&-a_{5}(x,\tau
)\quad \Rightarrow \quad a_{5\mathcal{T}}(x,\tau )=-a_{5}(x,-\tau )\quad .
\end{eqnarray}%
Under the combination of transformations 
\begin{eqnarray}
\psi (x,\tau )\mathrel{\mathop{\longrightarrow }\limits_{C}}\psi _{C}(x,\tau
) &=&\psi ^{\ast }(x,-\tau )  \label{c-1} \\
\tau \mathrel{\mathop{\longrightarrow }\limits_{C}}\tau _{C} &=&-\tau
\label{c-2} \\
a^{\mu }(x,\tau )\mathrel{\mathop{\longrightarrow }\limits_{C}}a_{C}^{\mu
}(x,\tau ) &=&a^{\mu }(x,-\tau )  \label{c-3} \\
a^{5}(x,\tau )\mathrel{\mathop{\longrightarrow }\limits_{C}}a_{C}^{5}(x,\tau
) &=&-a^{5}(x,-\tau )\quad ,  \label{c-4}
\end{eqnarray}%
if they can be made consistent with the pre-Maxwell equations, the charge
conjugate solution is $\psi _{C}(x,\tau )=\psi ^{\ast }(x,-\tau )$.

To check the consistency of the transformations (\ref{c-2}) to (\ref{c-4}),
we first find the 3-vector field strengths 
\begin{eqnarray}
e^{k} &=&f^{0k}=\partial ^{0}a^{k}-\partial ^{k}a^{0}\quad \mathrel{\mathop{
\longrightarrow }\limits_{C}}\quad e^{k} \\
h^{k} &=&\varepsilon ^{kij}\partial _{i}a_{j}\quad \mathrel{\mathop{
\longrightarrow }\limits_{C}}\quad h^{k} \\
\epsilon ^{k} &=&f^{5k}=\sigma \partial _{\tau }a^{k}-\partial
^{k}a_{5}\quad \mathrel{\mathop{\longrightarrow }\limits_{C}}\quad -\epsilon
^{k} \\
\epsilon ^{0} &=&f^{50}=\sigma \partial _{\tau }a^{0}-\partial
^{0}a_{5}\quad \mathrel{\mathop{\longrightarrow }\limits_{C}}\quad -\epsilon
^{0}\quad .
\end{eqnarray}%
We now consider the Lorentz force equations 
\begin{eqnarray}
M\frac{d^{2}x^{0}}{d\tau ^{2}} &=&e_{0}\left[ \mathbf{e}\cdot \frac{d\mathbf{%
x}}{d\tau }-\sigma \epsilon ^{0}\right] \\
M\frac{d^{2}\mathbf{x}}{d\tau ^{2}} &=&e_{0}\left[ \mathbf{e}\,\,\frac{dx^{0}%
}{d\tau }+\frac{d\mathbf{x}}{d\tau }\times \mathbf{h}-\sigma \mathbf{%
\epsilon }\right]
\end{eqnarray}%
which become 
\begin{eqnarray}
M\frac{d^{2}x_{C}^{0}}{d\tau ^{2}} &=&e_{0}\left[ \mathbf{e}_{C}\cdot \left( 
\frac{d\mathbf{x}}{d\tau }\right) _{C}-\sigma \epsilon _{C}^{0}\right] \\
M\frac{d^{2}\mathbf{x}_{C}}{d\tau ^{2}} &=&e_{0}\left[ \mathbf{e}_{C}\left( 
\frac{dx^{0}}{d\tau }\right) _{C}+\left( \frac{d\mathbf{x}}{d\tau }\right)
_{C}\times \mathbf{h}_{C}-\sigma \mathbf{\epsilon }_{C}\right]
\end{eqnarray}%
so that 
\begin{eqnarray}
M\frac{d^{2}x^{0}}{d\tau ^{2}} &=&e_{0}\left[ \mathbf{e}\cdot \left( -\frac{d%
\mathbf{x}}{d\tau }\right) -\sigma \left( -\epsilon ^{0}\right) \right] \\
M\frac{d^{2}\mathbf{x}}{d\tau ^{2}} &=&e_{0}\left[ \mathbf{e}\left( -\frac{%
dx^{0}}{d\tau }\right) +\left( -\frac{d\mathbf{x}}{d\tau }\right) \times 
\mathbf{h}-\sigma \left( -\mathbf{\epsilon }\right) \right]
\end{eqnarray}%
and finally 
\begin{eqnarray}
M\frac{d^{2}x^{0}}{d\tau ^{2}} &=&-e_{0}\left[ \mathbf{e}\cdot \frac{d%
\mathbf{x}}{d\tau }-\sigma \epsilon ^{0}\right]  \label{cl-Q-1} \\
M\frac{d^{2}\mathbf{x}}{d\tau ^{2}} &=&-e_{0}\left[ \mathbf{e}\frac{dx^{0}}{%
d\tau }+\frac{d\mathbf{x}}{d\tau }\times \mathbf{h}-\sigma \mathbf{\epsilon }%
\right] \quad \quad .  \label{cl-Q-2}
\end{eqnarray}%
Equations (\ref{cl-Q-1}) and (\ref{cl-Q-2}) are just the Lorentz equations (%
\ref{lor0-1}) and (\ref{lor2-1}) under the substitution $e_{0}\rightarrow
-e_{0}$. Thus, while form invariance under $\tau $-inversion is not a
reasonable symmetry to expect in pre-Maxwell theory and was not considered
along with space and time reversal symmetry, we find that the classical
operations 
\begin{equation}
\tau \rightarrow -\tau \quad \quad a_{5}\rightarrow -a_{5}
\end{equation}%
associated with quantum charge conjugation lead to a \emph{classical charge
conjugation} operation.

The action of this charge conjugation on the field equations is 
\begin{eqnarray}
\nabla \cdot \mathbf{e}-\left( -\partial _{\tau }\right) \left( -\epsilon
^{0}\right) &=&ej_{C}^{0} \\
\nabla \cdot \mathbf{e}-\partial _{\tau }\epsilon ^{0} &=&ej_{C}^{0}\quad
\Rightarrow \quad j_{C}^{0}=j^{0}
\end{eqnarray}%
\begin{eqnarray}
\nabla \times \mathbf{h}-\partial _{0}\mathbf{e}-\left( -\partial _{\tau
}\right) \left( -\mathbf{\epsilon }\right) &=&e\mathbf{j}_{C} \\
\nabla \times \mathbf{h}-\partial _{0}\mathbf{e}-\partial _{\tau }\mathbf{%
\epsilon } &=&e\mathbf{j}_{C}\quad \Rightarrow \quad \mathbf{j}_{C}=\mathbf{j%
}
\end{eqnarray}%
\begin{eqnarray}
\nabla \cdot \left( -\mathbf{\epsilon }\right) +\partial _{0}\left(
-\epsilon ^{0}\right) &=&ej_{C}^{5} \\
\nabla \cdot \mathbf{\epsilon }+\partial _{0}\epsilon ^{0}
&=&-ej_{C}^{5}\quad \Rightarrow \quad j_{C}^{5}=-j^{5}
\end{eqnarray}%
\begin{eqnarray}
\nabla \cdot \mathbf{h} &=&0 \\
\nabla \times \mathbf{e}+\partial _{0}\mathbf{h} &=&0
\end{eqnarray}%
\begin{eqnarray}
\nabla \times \left( -\mathbf{\epsilon }\right) -\sigma \left( -\partial
_{\tau }\right) \mathbf{h} &=&0 \\
\nabla \times \mathbf{\epsilon }-\sigma \partial _{\tau }\mathbf{h} &=&0
\end{eqnarray}%
\begin{eqnarray}
\nabla \left( -\epsilon ^{0}\right) +\sigma \left( -\partial _{\tau }\right) 
\mathbf{e}+\partial _{0}\left( -\mathbf{\epsilon }\right) &=&0 \\
\nabla \epsilon ^{0}+\sigma \partial _{\tau }\mathbf{e}+\partial _{0}\mathbf{%
\epsilon } &=&0
\end{eqnarray}%
and so we see that, taking the action on the currents as 
\begin{equation}
\left( j^{0},\mathbf{j},j^{5}\right) \mathrel{\mathop{\longrightarrow
}\limits_{C}}\left( j^{0},\mathbf{j},j^{5}\right) _{C}=\left( j^{0},\mathbf{j%
},-j^{5}\right) \quad ,
\end{equation}%
the pre-Maxwell equations transform consistently under the action of
classical charge conjugation. Moreover, we find that current conservation 
\begin{equation}
\partial _{\mu }j^{\mu }+\partial _{\tau }j^{5}=0\;\;\;\mathrel{\mathop{%
\longrightarrow }\limits_{C}}\;\;\;\partial _{\mu }j^{\mu }+\left( -\partial
_{\tau }\right) \left( -j^{5}\right) =0  \label{cur2}
\end{equation}%
is preserved. In quantum mechanics, the current $j^{5}$ is interpreted as
the probability of finding a particle in a localized volume of space time at
a given $\tau $, and the meaning of $j_{C}^{5}=-j^{5}$ must be examined
carefully.

\section{Conclusions}

The standard account of the discrete symmetries in quantum theory is deeply
influenced by Wigner's prescription \cite{movies} for time reversal, which
operates on both the coordinate time and the temporal ordering of events. It
should be noted that Wigner was concerned (both in 1932 and 1959) with the
non-relativistic quantum mechanics of atomic spectra, and his explicit use
of Galilean time determined his notion of time reversal\footnote{%
Wigner comments in \cite{movies}, ``Hence, `reversal of the direction of
motion' is perhaps a more felicitous, though longer, expression than `time
inversion.' ''}. Thus, neither negative energies nor pair
creation/annihilation played any part in his considerations. If, in the
spirit of Stueckelberg, we wish to disentangle the symmetries of the
coordinate time $t$ from those of the temporal parameter $\tau$, then we
expect that the discrete reflections will lead to the following
interpretations:

\begin{enumerate}
\item Space inversion covariance $P$ implies certain symmetric relations
between a given experiment and one performed in a spacially reversed
configuration.

\item Time inversion covariance $T$ implies certain symmetric relations
between a given experiment and one performed in a time-reversed
configuration, which is to say one in which advancement in $t$ is replaced
by retreat, and so a trajectory with $\dot x^0 > 0$ is replaced by a
trajectory with $\dot x^0 < 0$. Thus, we expect symmetric behavior between
pair annihilation processes and pair creation processes.

\item Charge conjugation covariance $C$ implies certain symmetric relations
between a given experiment and one in which the events are traced out in the
reverse order and carry opposite charge. Applying (\ref{670}) to this case,
we expect that the additive charges associated with internal symmetries also
undergo inversion.
\end{enumerate}

The analysis of the discrete symmetries in the Stueckelberg formalism ---
extended to include the parameter $\tau $ in local gauge group ---
demonstrates the theory's form invariance under spacetime inversion, and
exposes the charge conjugation symmetry, leading naturally to a view of
these symmetries based on interpretations 1 to 3 above. The structure of the
classical electromagnetic theory requires that, unlike the case for the
Maxwell field, the off-shell gauge field behave tensorially under the
discrete Lorentz transformations $P$ and $T$. Given these conditions on the
gauge fields, the quantum theory is seen to be invariant in a very simple
way under space and time reversal, and we may identify interpretations 1 and
2 --- the space inversion operation $P$ exchanges a particle trajectory with
its mirror image, and the time inversion operation $T$ exchanges particle
trajectories with antiparticle trajectories.

On the other hand, we do not regard the charge conjugation operation as
connecting symmetrical dynamical evolutions. The requirement that solutions
exist for the charge reversed case leads to an operation which reverses $%
x^{5}=\tau$ and the corresponding fifth-component objects, $a^{5}$ and $%
j^{5} $. Thus, the resulting charge conjugation operation, reverses the
temporal order of events and the sign of the charge and related currents,
leading to a negative probability density 
\begin{equation}
j^{5}=\Bigl|\psi (x,\tau )\Bigr|^{2} \mathrel{\mathop{\longrightarrow}%
\limits_{C}}j_{C}^{5}=-j^{5} \ \ ,  \label{neg-prob}
\end{equation}
which only makes sense in the context of the current conservation expression
(\ref{cur2}). Rather we associate the reversal of temporal order performed
by charge conjugation with the re-ordering of events performed by the
observer in the laboratory, who interprets events as always evolving from
earlier to later values of $t$. Thus, charge conjugation exchanges the
viewpoint of the events under interaction with the viewpoint of the
laboratory observer. The inversion of charges (associated with the gauge
symmetry and any internal symmetries) under this exchange reinforces the
conventional view of antiparticles in the laboratory, but does not influence
the event dynamics. Following Stueckelberg, we return to a formalism of
events interacting through gauge fields with events which may propagate
equivalently with $dt/d\tau <0$ or $dt/d\tau >0$, and understand the
antiparticle to simply be that part of an event trajectory for which $%
dt/d\tau <0$. The significance of the charge conjugation operation is that
the reversal of quantum numbers is observed in the laboratory when the
observer uses the laboratory clock as the parameter which orders the events.

In the context of Horwitz-Piron theory, the discrete symmetries of
(\ref{90}) were studied in \cite{concat}. That study, which assumed the standard
transformation properties for the four-vector Maxwell potential 
\begin{equation}
\left( A^{0},\mathbf{A}\right) \mathrel{\mathop{\rightarrow }\limits_{P}}%
\left( A^{0},-\mathbf{A}\right) 
\mbox{\qquad\qquad}%
\left( A^{0},\mathbf{A}\right) \mathrel{\mathop{\rightarrow }\limits_{T}}%
\left( A^{0},-\mathbf{A}\right) \;\;\;,
\end{equation}%
similarly concluded that while events interact without concern for the
particle/antiparticle distinction, the $CPT$ conjugate of the negative
energy trajectory is observed in the laboratory as the antiparticle.

A field theoretic study of the discrete symmetries and their significance in
quantum scattering will be reported in a forthcoming paper.


\end{document}